\documentclass[10pt,conference]{IEEEtran}

\IEEEoverridecommandlockouts
\usepackage[cmex10]{amsmath}
\usepackage{stfloats}
\usepackage{amssymb}
\usepackage{stackrel}

\usepackage{graphicx}
\usepackage{cite}
\usepackage{enumerate}

\newcommand{\Rmnum}[1]{\expandafter\@slowromancap\romannumeral #1@}

\begin{document}

\title{On the Secrecy Performance of NOMA  Systems with both External and Internal Eavesdroppers}

\author
{
Milad Abolpour, Mahtab Mirmohseni and Mohammad Reza Aref\\
Information Systems and Security Lab (ISSL) \\
Department of Electrical Engineering, Sharif University of Technology, Tehran, Iran\\
Email: miladabolpour@gmail.com,\{mirmohseni,aref\}@sharif.edu}

\maketitle
\begin{abstract}
Sharing resource blocks in NOMA systems provides more opportunity to the internal users to overhear the messages of the other users. Therefore, some sort of secrecy against the internal users in addition to the external eavesdroppers must be provided. In this paper, we investigate the secrecy performance of a two-user NOMA system in existence of the external and internal passive eavesdroppers, where the far user acts as an internal eavesdropper and tries to overhear the message of the near user.   Our system consists of a single antenna base station, two legitimate users and an external passive eavesdropper.  We  present the closed-forms for the ergodic secrecy rates of the users. Moreover, to derive the secrecy outage probability (SOP) of the system, we use Gaussian-Chebyshev quadrature method, which gives an approximation for the SOP. Numerical results show that this approximation is very close to the exact value of the SOP of the system. Finally, we eliminate the external eavesdropper and present the closed-forms for the ergodic rate of the far user, the ergodic secrecy rate of the near user and also the SOP of the system.

\end{abstract}
\section{Introduction}
Since non-orthogonal multiple access (NOMA) serves multiple users at the same time and frequency resources, using NOMA in fifth-generation (5G) networks enhances the spectral efficiency.  NOMA uses superposition at the base station \nolinebreak(BS) and successive interference cancelation (SIC) at the strong users \cite{islam2017power}.  The outage probability (OP)  of a randomly deployed users  NOMA  is studied in \cite{ding2014performance},  which  is shown that  NOMA enhances the outage performance of the system in comparison with orthogonal multiple access  systems. There exist many  works that investigate the OP of  the different types of NOMA systems such as cooperative NOMA in \cite{ding2015cooperative}, relay assisted NOMA in \cite{7925775,8490219} and NOMA with energy harvesting in \cite{ha2018outage,8555835}. Another system performance metric is the ergodic sum-rate, which is shown to be enhanced in a NOMA system in \cite{ding2014performance}. The ergodic rates of the users in different NOMA scenarios have been also studied in  \cite{7222419,8469484}.

As the messages are sent in the open medium to all users in  wireless networks, the secrecy of the messages of the users must be provided against the internal and external eavesdroppers. Utilizing the physical layer capabilities of wireless networks is a promising  way to enhance the secrecy performance of the system. Secrecy outage probability  (SOP) of a NOMA system with the external eavesdroppers is studied in \cite{liu2017enhancing,lei2017secure}. SOP of the other various  NOMA systems in existence of the external eavesdroppers have been investigated  in different scenarios such as  secure cooperative NOMA systems in \cite{chen2018physical,Abolpour2019seccoopNOMA,bassem2018passive} and  relay assisted NOMA networks  in \cite{8269229,8703400}. The ergodic secrecy rate of the users in  NOMA systems with the external eavesdropper is also investigated in \cite{8629032,8333686,yuan2019analysis}.

As users that share the same resource blocks in NOMA systems may not trust each other, at least some level of  secrecy must be provided against the internal users. In power domain NOMA systems, the  near users first carry out the SIC and decode the messages of the far users, then they decode their own messages. Therefore, they are aware of the messages of the far users, which means following the NOMA protocol forces the far users to trust the near users. But in some scenarios, it is necessary to maintain the secrecy at the near users against the far users. First in  \cite{bassem2018untrusted}, the secrecy performance of a NOMA system with an internal eavesdropper is investigated, in which  the SOP of  the near user and the OP of the system are derived.

In this paper, we study the effects of the existence of  both external and internal passive eavesdroppers on the secrecy performance of a NOMA system. We consider a system consisting of a single antenna base station, two legitimate users and an external passive eavesdropper. The users are called the near and far users according to their distances to the base station. The external eavesdropper is interested in overhearing the messages of both users. The far user trusts the near user in order to follow the power domain NOMA protocol, while it is interested in overhearing the message of the near user. We derive  the ergodic secrecy rate of the users and the SOP of the system in a closed form. Finally, by ignoring the  existence of the external eavesdropper, we reduce our system to the one in \cite{bassem2018untrusted} and derive   the ergodic secrecy rate of the near user and also the ergodic rate of the far user, not studied in \cite{bassem2018untrusted}. Moreover, we provide the closed-form  SOP of both users. Compared to the results of \cite{bassem2018untrusted}, which is the SOP of the near user and OP of both users, deriving the SOP of both users   is  more complex  due to the correlation between the channel coefficients. In presence of the external eavesdropper, we use Gaussian-Chebyshev quadrature method to  present an approximation for the SOP. Our simulations confirm the accuracy of this approximation.
\\
\textbf{\textit{Notation:}} In this paper, we denote the cumulative distribution function and probability density function of  a random variable $X$ as $F_{X}\left(x\right)$ and $f_{X}\left(x\right)$, respectively. $\overline{A}$ shows the complementary event of  event $A$ and $\text{Ei}\left(.\right)$ is the exponential integral, where $\text{Ei}\left(x\right)=-\int\limits_{x}^{\infty} \frac{e^{-t}}{t} \mathrm{d}t$. Moreover, $U(.)$ is the unit step function.

\section{System Model}
Our system consists of a single antenna BS, an  external passive eavesdropper, $U_{e}$, and  two legitimate users, $U_{m}$ and $U_{n}$, while it is assumed that $U_{m}$ is nearer than $U_{n}$ to the BS. The channel between $U_{e}$, $U_{m}$ and $U_{n}$ and BS are  rayleigh  fading channels with coefficients as $g_{e}$, $g_{m}$ and $g_{n}$, respectively. The $|g_{e}|^{2}$,$|g_{m}|^{2}$ and $|g_{n}|^{2}$ are   exponentially random variables  distributed with parameter $\lambda$. Also, $d_{e}$, $d_{m}$ and $d_{n}$ are the distances between $U_{e}$, $U_{m}$ and $U_{n}$ to the BS, respectively. Secrecy at both users are guaranteed against the $U_{e}$ and also secrecy at the near user is provided against the $U_{n}$.

 As depicted in  Fig. 1,  BS   with power  $P_{BS}$,  transmits a superposition of the messages of both users, called $S_{m}$ and $S_{n}$. The allocated power coefficients to  $U_{m}$ and $U_{n}$ are denoted as $a_{m}$ and $a_{n}$, respectively. So the transmitted signal of the BS is as:
\begin{equation}
\begin{aligned}
\label{eq: X_BS}
X_{BS}=\left(a_{m}S_{m}+a_{n}S_{n} \right)\sqrt{P_{BS}}.
\end{aligned}
\end{equation}
By following the NOMA protocol, we allocate more power to the far user, and thus $a_{n}^{2} \geq a_{m}^{2}$. At the receivers, $U_{m}$ and $U_{n}$ observe the signals as:
\begin{equation}
\begin{aligned}
\label{eq: Y_i}
Y_{i}=\frac{g_{i}\left(a_{m}S_{m}+a_{n}S_{n} \right)\sqrt{P_{BS}}}{d_{i}^{\frac{\alpha}{2}}}+N_{i},
\end{aligned}
\end{equation}
where $i \in \left\lbrace  U_{m},U_{n}  \right\rbrace$,  the path-loss is  denoted as $\alpha$ and  $N_{i}$ is a zero mean additive white Gaussian noise (AWGN) with variance one. $U_{m}$ decodes $S_{n}$ by $Y_{m}$ and then by subtracting it from $Y_{m}$ observes a signal as:
\begin{equation}
\begin{aligned}
\hat{Y}_{m}=\frac{g_{m}a_{m}\sqrt{P_{BS}}}{d_{m}^{\frac{\alpha}{2}}}S_{m}+N_{m}.
\end{aligned}
\end{equation}
 Simultaneously, $U_{n}$ decodes $S_{n}$ from $Y_{n}$, while it considers $S_{m}$ as noise.
\begin{figure}[t]
\centering
\includegraphics[width=9cm,height=4cm]{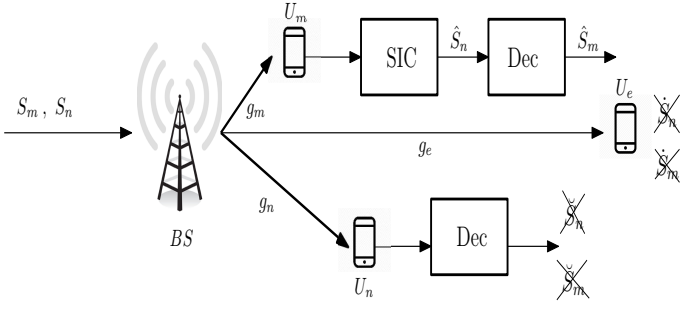}
\caption{System model with external eavesdropper.}
\end{figure}
\begin{figure}[t]
\centering
\includegraphics[width=9cm,height=4cm]{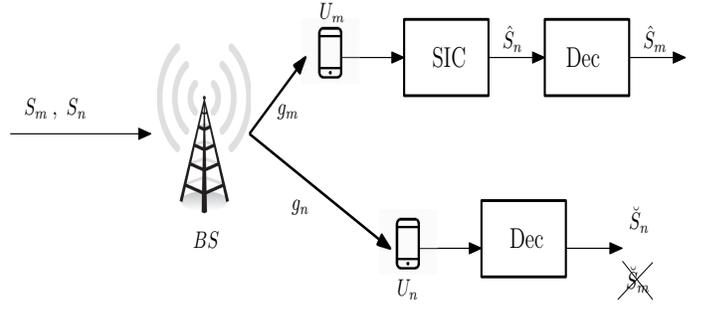}
\caption{System model without external eavesdropper.}
\end{figure}
The received signal at the external eavesdropper is as:
\begin{equation}
\begin{aligned}
Y_{e}=\frac{g_{e}\left(a_{m}S_{m}+a_{n}S_{n} \right)\sqrt{P_{BS}}}{d_{e}^{\frac{\alpha}{2}}}+N_{e},
\end{aligned}
\end{equation}
where  $N_{e}$ is the zero mean AWGN with variance one. $U_{e}$ first decodes $S_{n}$ from $Y_{e}$ and then decodes $S_{m}$ from  \linebreak $\hat{Y}_{e}=\frac{g_{e}a_{m}\sqrt{P_{BS}}}{d_{e}^{\frac{\alpha}{2}}}S_{m}+N_{e}$.

\subsection{No external eavesdropper}
As illustrated in Fig. 2,  by ignoring the external eavesdropper and assuming  $|g_{e}|^{2}=0$, the far user  overhears the message of the near user in a system as \cite{bassem2018untrusted}. The transmitted signal of the BS is as \eqref{eq: X_BS} and the received signals of the near and far users are as \eqref{eq: Y_i}. We note that in Fig. 1 and Fig. 2, $\hat{S}_{i}$, $\breve{S}_{i}$ and  $\dot{S}_{i}$ are denoting the decoded messages at $U_{m}$, $U_{n}$ and $U_{e}$, respectively, where $i \in \lbrace m,n \rbrace$.
\section{Secrecy Outage Probability}
The secrecy of the message of the far user must be provided against the $U_{e}$ and  also the secrecy of the message of the near user must be maintained against the $U_{e}$ and $U_{n}$. We assume $|\tilde{g}_{i}|^{2}=\frac{|g_{i}|^{2}}{d_{i}^{\alpha}}$, in which  $i \in \left\lbrace m,n,e       \right\rbrace$ and \linebreak $F_{|\tilde{g}_{i}|^{2}}\left( x\right)=\left(  1-e^{-\lambda d_{i}^{\alpha} x}   \right)$. By defining $R_{m}$ and $R_{n}$ as the targeted data rates of $U_{m}$ and $U_{n}$, the SOP event occurs if  either $U_{m}$ is not able to decode  $S_{n}$, i.e. ,
\begingroup\makeatletter\def\f@size{9}\check@mathfonts
\begin{equation}
\label{eq: E_1}
\overline{E}_{1}=\big\lbrace ( 1+\frac{|\tilde{g}_{m}|^{2}a_{n}^{2}}{|\tilde{g}_{m}|^{2}a_{m}^{2}+\frac{1}{P_{BS}}} )- \log ( 1+\frac{|\tilde{g}_{e}|^{2}a_{n}^{2}}{|\tilde{g}_{e}|^{2}a_{m}^{2}+\frac{1}{P_{BS}}}  )<R_{n}  \big\rbrace,
\end{equation}
\endgroup
or $U_{m}$ is not able to decode $S_{m}$, i.e. ,
\begingroup\makeatletter\def\f@size{9}\check@mathfonts
\begin{equation}
\begin{aligned}
\label{eq: E_2}
 &\overline{E}_{2}= \big\lbrace  \lbrace \log(  1+|\tilde{g}_{m}|^{2} a_{m}^{2}P_{BS} ) - \log(  1+|\tilde{g}_{n}|^{2} a_{m}^{2}P_{BS})<R_{m} \rbrace \   \\
 &   \cup \  \lbrace \log (  1+|\tilde{g}_{m}|^{2}a_{m}^{2}P_{BS} ) -  \log(  1+|\tilde{g}_{e}|^{2} a_{m}^{2}P_{BS}  )<R_{m}  \rbrace   \big\rbrace,
 \end{aligned}
\end{equation}
\endgroup
 or $U_{n}$ is not able to decode its own message, i.e. ,
 \begingroup\makeatletter\def\f@size{8.5}\check@mathfonts
\begin{equation}
\begin{aligned}
\label{eq: E_3}
\overline{E}_{3}=\big\lbrace   \log  (1+\frac{|\tilde{g}_{n}|^{2}a_{n}^{2}}{|\tilde{g}_{n}|^{2}a_{m}^{2}+\frac{1}{P_{BS}}}  )-  \log  (1+\frac{|\tilde{g}_{e}|^{2}a_{n}^{2}}{|\tilde{g}_{e}|^{2}a_{m}^{2}+\frac{1}{P_{BS}}}   )<R_{n}   \big\rbrace.
\end{aligned}
\end{equation}
\endgroup
Therefore, the SOP event is defined as:
\begin{equation}
\begin{aligned}
\label{eq: SOPdefinition}
\text{SOP}=1-\text{Pr}\left( E_{1} \cap E_{2} \cap E_{3}  \right).
\end{aligned}
\end{equation}
According to \eqref{eq: E_2} and \eqref{eq: E_3}, in order to avoid occurring secrecy outage, if we have $1+|\tilde{g}_{m}|^{2}a_{m}^{2}P_{BS} \geq 1+|\tilde{g}_{n}|^{2}a_{m}^{2}P_{BS}$ and $1+\frac{|\tilde{g}_{n}|^{2}a_{n}^{2}}{|\tilde{g}_{n}|^{2}a_{m}^{2}+\frac{1}{P_{BS}}} \geq 2^{R_{n}}\big(1+\frac{|\tilde{g}_{e}|^{2}a_{n}^{2}}{|\tilde{g}_{e}|^{2}a_{m}^{2}+\frac{1}{P_{BS}}}  \big)$, then we are sure that $1+\frac{|\tilde{g}_{m}|^{2}a_{n}^{2}}{|\tilde{g}_{m}|^{2}a_{m}^{2}+\frac{1}{P_{BS}}} \geq 2^{R_{n}}\big(1+\frac{|\tilde{g}_{e}|^{2}a_{n}^{2}}{|\tilde{g}_{e}|^{2}a_{m}^{2}+\frac{1}{P_{BS}}} \big)$, which means:
\begin{equation}
\begin{aligned}
\label{eq: intersection}
E_{1}\subseteq E_{2} \cap E_{3}.
\end{aligned}
\end{equation}
 Here, we rewrite the SOP as the summation of the probabilities of  two distinct events as:
\begin{equation}
\begin{aligned}
\text{SOP}=1-\left( r_{3}+r_{4}   \right),
\end{aligned}
\end{equation}
where $r_{3}$ and $r_{4}$ are defined at the bottom of this page in \eqref{eq: r_3} and \eqref{eq: r_4}, respectively.  According to  \eqref{eq: r_4},  $|\tilde{g}_{n}|^{2} < |\tilde{g}_{e}|^{2}$ and $1+\frac{|\tilde{g}_{n}|^{2}a_{n}^{2}}{|\tilde{g}_{n}|^{2}a_{m}^{2}+\frac{1}{P_{BS}}} \geq 2^{R_{n}}\left(1+\frac{|\tilde{g}_{e}|^{2}a_{n}^{2}}{|\tilde{g}_{e}|^{2}a_{m}^{2}+\frac{1}{P_{BS}}}  \right) $, therefore \linebreak$r_{4}=0$  and thus  $\text{SOP}=1-r_{3}$. The last step for deriving the SOP is calculating $r_{3}$. By defining $C_{1}=2^{R_{n}}-1$, \linebreak$X=\frac{|\tilde{g}_{e}|^{2}a_{n}^{2}}{|\tilde{g}_{e}|^{2}a_{m}^{2}+\frac{1}{P_{BS}}}$ and  $\delta_{1}=\frac{1-2^{R_{n}}a_{m}^{2}}{2^{R_{n}}a_{m}^{2}}$, we have:
\begin{figure*}[!b]
\hrulefill
\begingroup\makeatletter\def\f@size{8.75}\check@mathfonts
\begin{align}
\label{eq: r_3}
r_{3}= \text{Pr}  \left\lbrace  \left\lbrace 1+|\tilde{g}_{m}|^{2}a_{m}^{2}P_{BS} \geq 2^{R_{m}} \left(1+|\tilde{g}_{n}|^{2}a_{m}^{2}P_{BS} \right)     \right\rbrace \bigcap  \left\lbrace |\tilde{g}_{n}|^{2} \geq |\tilde{g}_{e}|^{2}\right\rbrace  \bigcap   \left\lbrace   1+\frac{|\tilde{g}_{n}|^{2}a_{n}^{2}}{|\tilde{g}_{n}|^{2}a_{m}^{2}+\frac{1}{P_{BS}}} \geq 2^{R_{n}}\left(1+\frac{|\tilde{g}_{e}|^{2}a_{n}^{2}}{|\tilde{g}_{e}|^{2}a_{m}^{2}+\frac{1}{P_{BS}}}  \right) \right\rbrace \right\rbrace.
\end{align}
\begin{align}
\label{eq: r_4}
r_{4}= \text{Pr} \left\lbrace  \left\lbrace 1+|\tilde{g}_{m}|^{2}a_{m}^{2}P_{BS} \geq 2^{R_{n}} \left( 1+|\tilde{g}_{e}|^{2}a_{m}^{2}P_{BS}  \right) \right\rbrace      \bigcap \lbrace |\tilde{g}_{n}|^{2}< |\tilde{g}_{e}|^{2} \rbrace \bigcap   \left\lbrace 1+\frac{|\tilde{g}_{n}|^{2}a_{n}^{2}}{|\tilde{g}_{n}|^{2}a_{m}^{2}+\frac{1}{P_{BS}}} \geq 2^{R_{n}}\left(1+\frac{|\tilde{g}_{e}|^{2}a_{n}^{2}}{|\tilde{g}_{e}|^{2}a_{m}^{2}+\frac{1}{P_{BS}}}  \right)\right\rbrace \right\rbrace.
\end{align}
\endgroup
\end{figure*}
\begin{equation}
\begin{aligned}
\label{eq: r_3p}
&r_{3}=\text{Pr} \big\lbrace    \lbrace  1+|\tilde{g}_{m}|^{2}a_{m}^{2}P_{BS} \geq 2^{R_{m}} \left(1+|\tilde{g}_{n}|^{2}a_{m}^{2}P_{BS} \right)\rbrace  \\
&   \cap \lbrace |\tilde{g}_{n}|^{2} \geq  \frac{C_{1}+2^{R_{n}}X}{P_{BS}\left(a_{n}^{2}-C_{1}a_{m}^{2}-2^{R_{n}}a_{m}^{2}X\right)} \rbrace \cap  \lbrace X\leq \delta _{1} \rbrace  \big\rbrace .
\end{aligned}
\end{equation}

By following \eqref{eq: r_3p}, we know that $X$ and $\delta_{1}$ are greater than zero, therefore $a_{m}^{2}$ must be lower than $\frac{1}{2^{R_{n}}}$, otherwise $r_{3}=0$ and SOP of the system becomes one. Now we rewrite \eqref{eq: r_3p} as:
\begingroup\makeatletter\def\f@size{10}\check@mathfonts
\begin{equation}
\begin{aligned}
\label{eq: pure r_{3}}
&r_{3}=\int\limits _{0}^{\delta_{1}}f_{X}\left(x \right) \int\limits_{ \frac{C_{1}+2^{R_{n}}x}{P_{BS}\left(a_{n}^{2}-C_{1}a_{m}^{2}-2^{R_{n}}a_{m}^{2}x\right)}}^{\infty}  f_{|\tilde{g}_{n}|^{2}}\left(y \right)   \times \\
&\int\limits_{\frac{2^{R_{m}}\left(1+a_{m}^{2}P_{BS}y\right)-1}{a_{m}^{2}P_{BS}}}^{\infty}     f_{|\tilde{g}_{m}|^{2}} \left(z \right)    \mathrm{d}z\mathrm{d}y \mathrm{d}x.
\vspace*{-2em}
\end{aligned}
\end{equation}
\endgroup
First we derive $F_{X} \left( x   \right)$ in order to calculate  $r_{3}$. So we have:
\begin{equation}
\begin{aligned}
\label{eq: F_X}
&F_{X}\left(x \right)=\text{Pr} \left\lbrace \frac{|\tilde{g}_{e}|^{2}a_{n}^{2}}{|\tilde{g}_{e}|^{2}a_{m}^{2}+\frac{1}{P_{BS}}}\leq x   \right\rbrace =U\left(x-\frac{a_{n}^{2}}{a_{m}^{2}}\right)+\\
&F_{|\tilde{g}_{e}|^{2}}\left(\frac{x}{P_{BS}\left( a_{n}^{2}-a_{m}^{2}x   \right)} \right)U\left(-x+\frac{a_{n}^{2}}{a_{m}^{2}} \right) =U\left(x-\frac{a_{n}^{2}}{a_{m}^{2}}\right)\\
&+\left(1-e^{-\lambda d_{e}^{\alpha }\frac{x}{P_{BS}\left( a_{n}^{2}-a_{m}^{2}x   \right)} } \right)U\left(-x+\frac{a_{n}^{2}}{a_{m}^{2}} \right).
\end{aligned}
\end{equation}
By taking derivative of  \eqref{eq: F_X}, $f_{X}\left( x\right)$ is obtained as:
\begin{equation}
\begin{aligned}
\label{eq: f_X}
&f_{X}\left( x\right)=\frac{\lambda d_{e}^{\alpha}}{P_{BS}}\frac{a_{n}^{2}}{\left(a_{n}^{2}-a_{m}^{2}x\right)^{2}} e^{-\lambda d_{e}^{\alpha }\frac{x}{P_{BS}\left( a_{n}^{2}-a_{m}^{2}x   \right)}}  \times  \\
&U\left(-x+\frac{a_{n}^{2}}{a_{m}^{2}} \right).
\end{aligned}
\end{equation}
By substituting \eqref{eq: f_X} into   \eqref{eq: pure r_{3}} and after some mathematical calculations, $r_{3}$ is obtained as:
\begingroup\makeatletter\def\f@size{8.5}\check@mathfonts
\begin{equation}
\begin{aligned}
\label{eq: r_3F}
&r_{3}= \frac{\lambda \left(d_{e}d_{n}\right)^{\alpha}}{P_{BS}\left( d_{n}^{\alpha}+2^{R_{m}}d_{m}^{\alpha}     \right)}e^{-\lambda \frac{\left(2^{R_{m}}-1\right)d_{m}^{\alpha}}{a_{m}^{2}P_{BS}}} \times \\
&\int\limits_{0}^{\delta _{m}}   \frac{a_{n}^{2}}{\left(a_{n}^{2}-a_{m}^{2}x\right)^{2}}e^{-\lambda \left(   d_{e}^{\alpha }\frac{x}{P_{BS}\left( a_{n}^{2}-a_{m}^{2}x   \right)}+  \frac{\left(C_{1}+2^{R_{n}}x\right)\left(  d_{n}^{\alpha}+2^{R_{m}}d_{m}^{\alpha}  \right)}{P_{BS}\left(a_{n}^{2}-C_{1}a_{m}^{2}-2^{R_{n}}a_{m}^{2}x\right)}      \right)}  \mathrm{d}x\\
&= \frac{\lambda \left(d_{e}d_{n}\right)^{\alpha}}{P_{BS}\left( d_{n}^{\alpha}+2^{R_{m}}d_{m}^{\alpha}     \right)}e^{-\lambda \frac{\left(2^{R_{m}}-1\right)d_{m}^{\alpha}}{a_{m}^{2}P_{BS}}} \times I,
\end{aligned}
\end{equation}
\endgroup
where $\delta_{m}= \min \left(\frac{a_{n}^{2}}{a_{m}^{2}},\delta_{1}   \right)$. We use Gaussian-Chebyshev quadrature method to approximate  $I$ as:
\begin{equation}
\begin{aligned}
\label{eq: I}
I\approx \sum\limits_{i=1}^{N} \omega _{i}e^{-\lambda c_{i}},
\end{aligned}
\end{equation}
where\begin{scriptsize}
$c_{i}=\frac{d_{e}^{\alpha }\left(  1+\theta _{i}  \right)\frac{\delta_{m}}{2}}{P_{BS}\left( a_{n}^{2}-a_{m}^{2}  \frac{\delta_{m}}{2}   \left(  1+\theta_{i} \right)   \right)}+$ \nolinebreak$\frac{\left(C_{1}+2^{R_{n}}\frac{\delta_{m}}{2}\left(1+\theta_{i}\right)\right)   \left(   d_{n}^{\alpha}+2^{R_{m}}d_{m}^{\alpha}   \right)  }{P_{BS}\left(a_{n}^{2}-C_{1}a_{m}^{2}-2^{R_{n}}a_{m}^{2}\frac{\delta_{m}}{2}\left(  1+\theta _{i}   \right)\right)}  $,
\end{scriptsize} $\omega _{i}= \frac{\pi \delta_{m}}{2N}\frac{a_{n}^{2}\sqrt{1-\theta_{i}^{2}} }{\left(a_{n}^{2}-a_{m}^{2} \frac{\delta_{m}}{2}   \left(  1+\theta _{i} \right)\right)^{2}}  $, $\theta_{i}= \cos \left(  \frac{2i-1}{2N} \pi  \right) $ and $N$ is the complexity-vs-accuracy coefficient. Finally by substituting \eqref{eq: I} into  \eqref{eq: r_3F} the SOP of the system is obtained as:
\begin{equation}
\begin{aligned}
\text{SOP}  \approx   1-  \frac{\lambda \left(d_{e}d_{n}\right)^{\alpha}}{P_{BS}\left( d_{n}^{\alpha}+2^{R_{m}}d_{m}^{\alpha}     \right)}e^{-\lambda \frac{\left(2^{R_{m}}-1\right)d_{m}^{\alpha}}{a_{m}^{2}P_{BS}}} \sum\limits_{i=1}^{N} \omega _{i}e^{-\lambda c_{i}}.
\end{aligned}
\end{equation}
\vspace*{-2em}
\subsection{No external eavesdropper}
By assuming  $|g_{e}|^{2}=0$, we reduce our system to a system  in \cite{bassem2018untrusted}. We consider a more general definition on the SOP, which includes  the secrecy outage probability of $U_{m}$ and total outage probability of the system, considered in \cite{bassem2018untrusted}.

The  SOP of the system is as \eqref{eq: SOPdefinition}, where due to the eliminating the eavesdropper, the event that $U_{m}$ is not able to decode $S_{n}$ and also the event that $U_{n}$ is not able to decode its own message are changed to  $\overline{E}_{1}=\left\lbrace     \log \left(1+\frac{|\tilde{g}_{m}|^{2}a_{n}^{2}}{|\tilde{g}_{m}|^{2}a_{m}^{2}+\frac{1}{P_{BS}}} \right)   < R_{n} \right\rbrace$  and $\overline{E}_{3}= \left\lbrace   \log \left(   1+  \frac{|\tilde{g}_{n}|^{2}a_{n}^{2}}{|\tilde{g}_{n}|^{2}a_{m}^{2}+\frac{1}{P_{BS}}}   \right)   < R_{n}   \right\rbrace   $, respectively. By following the same approach as \eqref{eq: intersection}, the SOP of the system is written as:

\begin{equation}
\begin{aligned}
\text{SOP}=\text{Pr} \left( E_{2} \ \cap  \ E_{3}     \right).
\end{aligned}
\end{equation}
By defining $\tau_{n}=\frac{C_{1}}{P_{BS}\left(  a_{n}^{2}-C_{1}a_{m}^{2}  \right)}$, after some mathematical calculations, SOP of the system becomes one under the condition $ \frac{a_{n}^{2}}{a_{m}^{2}} \leq  C_{1}$, otherwise the SOP of the system equals to:
\begin{equation}
\begin{aligned}
\label{eq: SOP_nearbrief}
\text{SOP}=1-P_{1},
\end{aligned}
\end{equation}
where $P_{1}=\text{Pr}   \left\lbrace   \lbrace |\tilde{g}_{n}|^{2} \geq \tau_{n}\rbrace  \cap \  \lbrace 1+|\tilde{g}_{m}|^{2} a_{m}^{2}P_{BS} \geq    \right. $ $ \left.   2^{R_{m}} \left(  1+|\tilde{g}_{n}|^{2} a_{m}^{2} P_{BS}    \right)  \rbrace \right\rbrace $, and is equal to:
\begin{equation}
\begin{aligned}
\label{eq: P_1int}
P_{1}=\int\limits_{\tau_{n}} ^{\infty} f_{|\tilde{g}_{n}|^{2}}\left( x \right) \int\limits_{\frac{2^{R_{m}}\left(1+a_{m}^{2}P_{BS}x\right)-1}{a_{m}^{2}P_{BS}}} ^{\infty} f_{|\tilde{g}_{m}|^{2}}\left( y \right) \mathrm{d}y\mathrm{d}x.
\end{aligned}
\end{equation}
By applying $f_{|\tilde{g}_{i}|^{2}}(x)=\lambda d_{i}^{\alpha} e^{-\lambda d_{i}^{\alpha}x} \ (i \in \lbrace m,n,e \rbrace)$  into \eqref{eq: P_1int}, $P_{1}$ is obtained as:
\begin{equation}
\begin{aligned}
\label{eq: P_1Final}
P_{1}=\frac{d_{n}^{\alpha}}{d_{n}+2^{R_{m}}d_{m}^{\alpha}} e^{-\lambda   \left(  \frac{d_{m}^{\alpha}\left( 2^{R_{m}}-1  \right)  }   {a_{m}^{2}P_{BS}} +    \tau_{n}   \left(  d_{n}^{\alpha} +2^{R_{m}}d_{m}^{\alpha} \right)    \right) }.
\end{aligned}
\end{equation}
Therefore, from \eqref{eq: SOP_nearbrief} and \eqref{eq: P_1Final}, the SOP of the system equals to:
\begin{equation}
\begin{aligned}
\text{SOP}=1-\frac{d_{n}^{\alpha}}{d_{n}+2^{R_{m}}d_{m}^{\alpha}} e^{-\lambda   \left(  \frac{d_{m}^{\alpha}\left( 2^{R_{m}}-1  \right)  }   {a_{m}^{2}P_{BS}} +    \tau_{n}   \left(  d_{n}^{\alpha} +2^{R_{m}}d_{m}^{\alpha} \right)    \right) }.
\end{aligned}
\end{equation}

\vspace*{-1.5em}
\section{Ergodic Secrecy Rate}
In this section, we study the ergodic secrecy rates of the users. First we derive the achievable rate of  each user without secrecy constraints and then we subtract  the leakage rate  from its achievable rate to obtain secure achievable rate, as:

\begin{equation}
\begin{aligned}
R_{i}^{sec}=R_{i}^{a}-R_{i}^{L},
\end{aligned}
\end{equation}
where  $R_{i}^{a}$ and $R_{i}^{L}$ are the  achievable and leakage rates of the user $i$, $ i \in \left\lbrace    U_{m},U_{n}   \right\rbrace $. Finally, the expected value  of $R_{i}^{sec}$ presents the ergodic secrecy rate of  the  user $i$, as:
\begin{equation}
\begin{aligned}
\label{eq: averageR}
\mathbb{E}(R_{i}^{sec})=\mathbb{E}(R_{i}^{a})-\mathbb{E}(R_{i}^{L}).
\end{aligned}
\end{equation}

Since the message of $U_{m}$ is overheard by $U_{n}$ and $U_{e}$, so the rate of the leakage information of $U_{m}$ is as:
\begin{equation}
\begin{aligned}
\label{eq: R_mbothL}
R_{m}^{L}&=\max \left(  1+|\tilde{g}_{n}|^{2}a_{m}^{2}P_{BS} \, , \, 1+|\tilde{g}_{e}|^{2}a_{m}^{2}P_{BS}   \right)\\
&=1+a_{m}^{2}P_{BS} \max \left( |\tilde{g}_{n}|^{2} \, , \, |\tilde{g}_{e}|^{2}    \right).
\end{aligned}
\end{equation}
Now, we find  $F_{\max \left( |\tilde{g}_{n}|^{2} \, , \, |\tilde{g}_{e}|^{2}    \right)} \left(  x  \right)$ in order to derive the  expected value of \eqref{eq: R_mbothL} as:
\begin{equation}
\begin{aligned}
\label{eq: F_max}
F_{\max \left( |\tilde{g}_{n}|^{2} \, , \, |\tilde{g}_{e}|^{2}    \right)} \left(  x  \right)&=
\text{Pr} \left\lbrace   |\tilde{g}_{n}|^{2} < x \cap  |\tilde{g}_{e}|^{2} <x   \right\rbrace \\
&=\left(  1-e^{-\lambda d_{n}^{\alpha}x}  \right)\left(  1-e^{-\lambda d_{e}^{\alpha}x}  \right).
\end{aligned}
\end{equation}
 By taking derivative  of  \eqref{eq: F_max}, $f_{\max \left( |\tilde{g}_{n}|^{2} \, , \, |\tilde{g}_{e}|^{2}    \right)} \left(  x  \right)$  equals to:
 \begin{equation}
 \begin{aligned}
& f_{\max \left( |\tilde{g}_{n}|^{2} \, , \, |\tilde{g}_{e}|^{2}    \right)} \left(  x  \right)=\\
 &\lambda \left(    d_{n}^{\alpha} e^{-  \lambda d_{n}^{\alpha}x}+  d_{e}^{\alpha} e^{  -\lambda    d_{e}^{\alpha}x} -\left( d_{n}^{\alpha}+d_{e}^{\alpha}   \right)  e^{-\lambda \left(  d_{n}^{\alpha}+d_{e}^{\alpha}  \right)x}  \right).
 \end{aligned}
 \end{equation}
  Now $\mathbb{E}(R_{m}^{L})$ is obtained as:
 \begingroup\makeatletter\def\f@size{9}\check@mathfonts
 \begin{equation}
 \begin{aligned}
 \label{eq: R_mbothlfinal}
&\mathbb{E}(R_{m}^{L})=\mathbb{E} \left\lbrace    1+a_{m}^{2}P_{BS} \max \left( |\tilde{g}_{n}|^{2} \, , \, |\tilde{g}_{e}|^{2}        \right)   \right\rbrace =\\
 & \left( \log _{2}e\right)  \int\limits _{0}^{\infty} \ln \left( 1+a_{m}^{2}P_{BS} x  \right)  f_{ \max \left( |\tilde{g}_{n}|^{2} \, , \, |\tilde{g}_{e}|^{2}    \right)      }\left( x \right) \mathrm{d}x =\\
 & -\left( \log _{2}e\right)   \left(  e^{\frac{\lambda d_{n}^{\alpha}}{a_{m}^{2}P_{BS}}}     \text{Ei}   \left( -   \frac{\lambda d_{n}^{\alpha}}{a_{m}^{2}P_{BS}}    \right)     + \right. \\
 &\left.     e^{\frac{\lambda d_{e}^{\alpha}}{a_{m}^{2}P_{BS}}}     \text{Ei}   \left( -   \frac{\lambda d_{e}^{\alpha}}{a_{m}^{2}P_{BS}}    \right)          - e^{\frac{\lambda \left(  d_{n}^{\alpha}+d_{e}^{\alpha}  \right)^{\alpha}}{a_{m}^{2}P_{BS}}}     \text{Ei}   \left( -   \frac{\lambda \left(  d_{n}^{\alpha}+d_{e}^{\alpha}  \right)^{\alpha}}{a_{m}^{2}P_{BS}}    \right) \right).
 \end{aligned}
 \end{equation}
 \endgroup

 The   achievable rate of $U_{m}$ is as:
 \begin{equation}
\begin{aligned}
R_{m}^{a}=\log \left(  1+|\tilde{g}_{m}|^{2}a_{m}^{2}P_{BS}  \right).
\end{aligned}
\end{equation}
 By following the approach of equations (27) and  (28) in \cite{yuan2019analysis}, and after some mathematical calculations, the expected value of  achievable rate of $U_{m}$ equals to:
 \begin{equation}
 \begin{aligned}
 \label{eq: averageRm}
  \mathbb{E}(R_{m}^{a})=-\log _{2}\left( e    \right)   e^{\frac{\lambda d_{m}^{\alpha}}{a_{m}^{2}P_{BS}}}   \text{Ei} \left(  -\frac{\lambda d_{m}^{\alpha}}{a_{m}^{2}P_{BS}} \right).
 \end{aligned}
 \end{equation}
Finally, by substituting \eqref{eq: R_mbothlfinal} and \eqref{eq: averageRm} into  \eqref{eq: averageR}, the ergodic secrecy rate of $U_{m}$ is obtained as:
\begin{equation}
\begin{aligned}
\label{eq: R_nbothLaverage}
&\mathbb{E}(R_{m}^{sec})=-\left(   \log_{2}e  \right) \left(e^{\frac{\lambda d_{m}^{\alpha}}{a_{m}^{2}P_{BS}}}    \text{Ei} \left(  -\frac{\lambda d_{m}^{\alpha}}{a_{m}^{2}P_{BS}}   \right)- \right. \\
& \left.    e^{\frac{\lambda d_{n}^{\alpha}}{a_{m}^{2}P_{BS}}}     \text{Ei}   \left( -   \frac{\lambda d_{n}^{\alpha}}{a_{m}^{2}P_{BS}}    \right)    -  e^{\frac{\lambda d_{e}^{\alpha}}{a_{m}^{2}P_{BS}}}     \text{Ei}   \left( -   \frac{\lambda d_{e}^{\alpha}}{a_{m}^{2}P_{BS}}    \right)  \right. \\
& \left. + e^{\frac{\lambda \left(  d_{n}^{\alpha}+d_{e}^{\alpha}  \right)^{\alpha}}{a_{m}^{2}P_{BS}}}     \text{Ei}   \left( -   \frac{\lambda \left(  d_{n}^{\alpha}+d_{e}^{\alpha}  \right)^{\alpha}}{a_{m}^{2}P_{BS}}    \right) \right) .
\end{aligned}
\end{equation}
The  leakage rate of $U_{n}$ is:
 \begin{equation}
 \begin{aligned}
 \label{eq: R_nbothL}
 R_{n}^{L}= \log \left(   1+\frac{|\tilde{g}_{e}|^{2}a_{n}^{2}P_{BS}} {|\tilde{g}_{e}|^{2}a_{n}^{2}P_{BS}+1}  \right) .
 \end{aligned}
 \end{equation}
 Therefore by following the approach of  \cite{yuan2019analysis}, the expected value of the leakage rate of the far user is obtained as:
 \begin{equation}
 \begin{aligned}
& \mathbb{E}(R_{n}^{L})  =\mathbb{E} \left\lbrace   \log \left(  1+P_{BS}|\tilde{g}_{e}|^{2}  \right) -\log \left(  1+a_{m}^ {2}P_{BS} | \tilde{g } _{e} |^{2} \right)  \right\rbrace = \\
& -  \left(  \log _{2}e \right) \left(  e^ {\frac{  \lambda d_{e}^{\alpha}}{P_{BS}} }  \text{Ei}\left(  -\frac{\lambda   d_{e}^{\alpha}}{P_{BS}}  \right) -e^{  \frac{ \lambda d_{e}^{\alpha }}{a_{m}^{2}P_{BS}}} \text{Ei}  \left(-  \frac{\lambda d_{e}^{\alpha}}{a_{m} ^{2}P_{BS}}  \right)  \right) .
 \end{aligned}
 \end{equation}
 The  achievable rate of $U_{n}$ is the minimum of the  achievable rate of the near user for performing SIC and  the  achievable  rate of the far user in order to decode its own message, which means:
\begin{equation}
\begin{aligned}
&R_{n}^{a}=\min \Big( \log \big( 1+\frac{|\tilde{g}_{m}|^{2}a_{m}^{2}}{|\tilde{g}_{m}|^{2}a_{m}^{2}+\frac{1}{P_{BS}}}  \big)  , \\
&  \log \big(  1+ \frac{|\tilde{g}_{n}|^{2}a_{n}^{2}}{|\tilde{g}_{n}|^{2}a_{m}^{2}+\frac{1}{P_{BS}}} \big)   \Big).
\end{aligned}
\end{equation}
The ergodic rate of $U_{n}$ is the expected value of $R_{n}^{a}$, which is shown in the following:
\begin{equation}
\begin{aligned}
\label{eq: R_nbar}
&\mathbb{E}(R_{n}^{a})=\mathbb{E} \left\lbrace   \log \left(  \min \left(  |\tilde{g}_{m}|^{2}, |\tilde{g}_{n}|^{2} \right)P_{BS}+1  \right) -   \right.    \\
& \left.   \log \left( \min  \left(  |\tilde{g}_{m}|^{2},|\tilde{g}_{n}|^{2}   \right) a_{m}^{2}P_{BS}+1  \right)    \right\rbrace.
\end{aligned}
\end{equation}
In order to calculate \eqref{eq: R_nbar}, we  derive  $F_{ \min \left(  |\tilde{g}_{m}|^{2}, |\tilde{g}_{n}|^{2} \right)} \left(x \right)$. So we have:
\begin{equation}
\begin{aligned}
F_{ \min \left(  |\tilde{g}_{m}|^{2}, |\tilde{g}_{n}|^{2} \right)} \left(x \right)&= \text{Pr} \left\lbrace  \min \left( |\tilde{g}_{m}|^{2}, |\tilde{g}_{n}|^{2}    \right) \leq x \right\rbrace  \\
 &\stackrel{(a)}{=}1-e^{-\lambda d x},
\end{aligned}
\end{equation}
where (a) holds due to the independence of the channel coefficients and also we define $d$ as $d=d_{m}^{\alpha} +d_{n}^{\alpha}$. So by taking derivative we have:
\begin{equation}
\begin{aligned}
\label{eq: f_min}
f_{ \min \left(  |\tilde{g}_{m}|^{2}, |\tilde{g}_{n}|^{2} \right)} \left(x \right)=
\lambda d e^{-\lambda d x}.
\end{aligned}
\end{equation}
After applying \eqref{eq: R_nbar} and \eqref{eq: f_min} and  some mathematical calculations and  following the approach of \cite{yuan2019analysis}, the ergodic rate of $U_{n}$ is derived as:
\begin{equation}
\begin{aligned}
\label{eq: R_n^a}
&\mathbb{E}(R_{n}^{a}) =- \Big(\log_{2}e\big) \times \\
&\big(e^{\frac{\lambda d}{P_{BS}}}\text{Ei}\big(-\frac{\lambda d}{P_{BS}} \big)  -  e^{\frac{\lambda d}{a_{m}^{2}P_{BS}}}\text{Ei}\big(-\frac{\lambda d}{a_{m}^{2}P_{BS}} \big)  \Big).
\end{aligned}
\end{equation}
Now by substituting \eqref{eq: R_nbar} and \eqref{eq: R_n^a} into \eqref{eq: averageR}, the ergodic secrecy rate of  $U_{n}$ is obtained as:
\begin{equation}
\begin{aligned}
&\mathbb{E}(R_{n}^{sec})=-  \left( \log_{2}e\right) \left(   \left(e^{\frac{\lambda d}{P_{BS}}}\text{Ei}\left(-\frac{\lambda d}{P_{BS}} \right)  -  \right. \right. \\ &  \left.  \left. e^{\frac{\lambda d}{a_{m}^{2}P_{BS}}}\text{Ei}\left(-\frac{\lambda d}{a_{m}^{2}P_{BS}} \right)  \right) -  e^ {\frac{  \lambda d_{e}^{\alpha}}{P_{BS}} }  \text{Ei}\left(  -\frac{\lambda   d_{e}^{\alpha}}{P_{BS}}  \right)+ \right. \\
& \left.e^{  \frac{ \lambda d_{e}^{\alpha }}{a_{m}^{2}P_{BS}}} \text{Ei}  \left(-  \frac{\lambda d_{e}^{\alpha}}{a_{m} ^{2}P_{BS}}  \right)   \right).
\end{aligned}
\end{equation}
\subsection{No external eavesdropper}
In this subsection, the existence of the eavesdropper is ignored by  assuming $|g_{e}|^{2}=0$. The ergodic rate of the far user and  the ergodic secrecy rate of the near user are provided, while the far user acts as an internal eavesdropper. The ergodic rate of $U_{n}$ equals to $\mathbb{E}(R_{n}^{a})$, which is derived   in \eqref{eq: R_n^a}. Moreover, by following the same way as \eqref{eq: averageRm}, the expected value of the leakage rate of $U_{m}$ equals to $- (\log_{2}e)  e^{\frac{\lambda d_{n}^{\alpha}}{a_{m}^{2}P_{BS}}}  \text{Ei} ( -   \frac{\lambda d_{n}^{\alpha}}{a_{m}^{2}P_{BS}} ) $. Therefore, by using \eqref{eq: averageRm}, the ergodic secrecy rate of \hspace*{.3em} $U_{m}$ is:
\begin{equation}
\begin{aligned}
\label{eq: R_n}
&\mathbb{E}(R_{m}^{sec})=-   \big(  \log_{2}e\big)  \times \\
& \Big(     e^{\frac{\lambda d_{m}^{\alpha}}{a_{m}^{2}P_{BS}}}  \text{Ei} \big( -   \frac{\lambda d_{m}^{\alpha}}{a_{m}^{2}P_{BS}}  \big)  - e^{\frac{\lambda d_{n}^{\alpha}}{a_{m}^{2}P_{BS}}}  \text{Ei} \big( -   \frac{\lambda d_{n}^{\alpha}}{a_{m}^{2}P_{BS}}  \big)   \Big).
\end{aligned}
\end{equation}

\section{Numerical Results}
In this section, we present the numerical results for two cases of   the  secrecy  outage probability and ergodic secrecy rates of the users. Parameters of our simulations are presented in Table \ref{table: Parameters}. Monte-Carlo simulations are utilized by generating random fading channel coefficients for the users and the external eavesdropper.
\begin{table}[tb]
\centering
\caption{Simulations  Parameters}
 \label{table: Parameters}
\begin{tabular}{|l|l|}
\hline
Number of iterations & $10^{5}$  \\
\hline
Distance of $U_{m}$ to BS& $d_{m}=4m$  \\
\hline
Distance of $U_{n}$ to BS & $d_{n}=6m$ \\
\hline
Distance of $U_{e}$ to BS & $d_{e}=7m$  \\
\hline
Parameter of the channels & $\lambda=1$  \\
\hline
Path-loss exponent & $\alpha=4$  \\
\hline
Users' targeted data rates  (Fig. 3 and Fig. 4)& $R_{n}=R_{m}=0.1$  \\
\hline
 Power allocation coefficients  (Fig. 3 and Fig. 5)& $a_{n}^{2}=0.6 , a_{m}^{2}=0.4$  \\
\hline
The complexity-vs-accuracy coefficient & $N=200$  \\
\hline
Transmitted power of $BS$ (Fig. 4. and Fig. 6.) & $P_{BS}=30$dB  \\
\hline
\end{tabular}
\end{table}

\begin{figure}[tb]
\includegraphics[width=8cm,height=5cm]{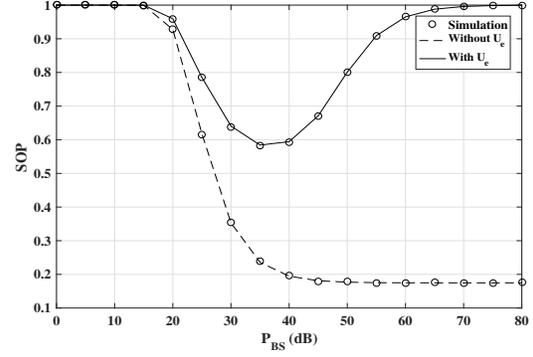}

\caption{SOP of the system versus $P_{BS}$.}
\end{figure}

\begin{figure}[tb]
\includegraphics[width=8cm,height=5cm]{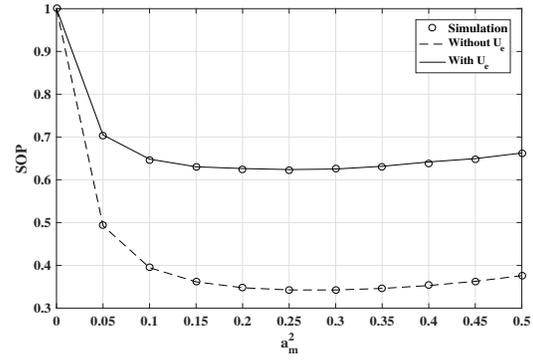}
\caption{SOP  of the system versus $a_{m}^{2}$.}
\vspace*{-2em}
\end{figure}

\begin{figure}[tb]
\includegraphics[width=8cm,height=5cm]{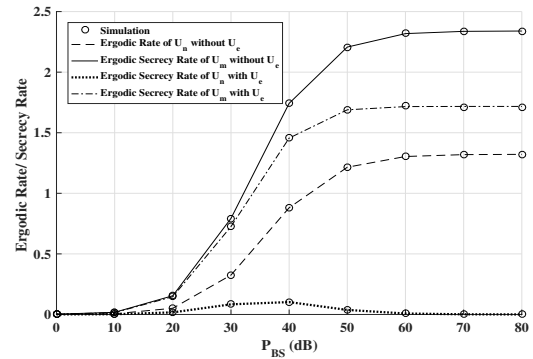}
\caption{Ergodic rate/ Secrecy rate of the users versus $P_{BS}$.}
\vspace*{-2em}
\end{figure}

\begin{figure}[tb]
\includegraphics[width=8cm,height=5cm]{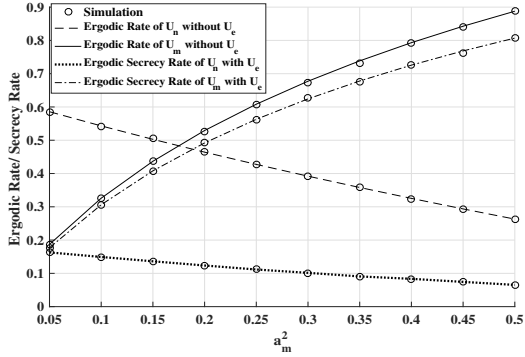}
\caption{Ergodic rate/ secrecy rate of the users versus $a_{m}^{2}$.}
\vspace*{-2em}
\end{figure}

\subsection{Secrecy outage probability}
Fig. 3 shows the SOP of the NOMA system with and without the $U_{e}$. As illustrated, the analytical results are confirmed by the simulation results. Without $U_{e}$ at low SNR regimes, increasing $P_{BS}$ decreases the SOP of the system. While at high SNR regimes, increasing the transmitted power of the base station has no effects on the SOP of the system, because the SOP only depends on the channel coefficients. In existence of the $U_{e}$, increasing the transmitted power of the base station at low SNR regimes decreases the SOP of the system. But at high SNR regimes due to the increment of the ability of the $U_{e}$ in order to decode the messages of $U_{m}$ and $U_{n}$, increasing $P_{BS}$ increases the SOP of the system.

Fig. 4 shows the SOP of the systems with and without $U_{e}$ with respect to $a_{m}^{2}$ (power allocation coefficient of NOMA). As mentioned, following the NOMA protocol forces us to allocate more power to the far user, which means $a_{m}^{2} \leq  0.5$. In both cases, we see that by increasing $a_{m}^{2}$ up to 0.25, SOP of the system decreases. But according to the increment of the ability of the eavesdroppers in order to decode the messages of the users and also decrement of the ability of $U_{n}$ in order to decode its own message, increasing $a_{m}^{2}$ from 0.25 to 0.5 decreases the SOP of the system. Moreover, existence of $U_{e}$ degrades the secrecy outage performance of the system.
\subsection{Ergodic secrecy rate}
Fig. 5 demonstrates the ergodic rate and ergodic secrecy rate of the users with respect to the transmitted power of the base station with and  without  $U_{e}$. We see that  without $U_{e}$, at low SNR regimes,  increasing $P_{BS}$ increases the ergodic rate of $U_{n}$. At high SNR regimes, i.e. ,  $P_{BS}\rightarrow \infty$, the ergodic rate of $U_{n}$ goes to $\log \left(1+\frac{a_{n}^{2}}{a_{m}^{2}}\right)=\log \left(2.5\right)=1.3219$, which is confirmed by our simulations. Moreover with $U_{e}$, increasing $P_{BS}$ increases the ergodic secrecy rate of $U_{n}$ at low SNR regimes.  Since the ergodic rates of  $U_{e}$ and $U_{n}$ goes to 1.3219 at high SNR regimes,  for the value of $P_{BS}$ higher than 40dB, increasing the transmitted power of the base station decreases the ergodic secrecy rate of $U_{n}$ until it goes to zero. Besides, with and without $U_{e}$, at low SNR regimes, increasing $P_{BS}$ increases the ergodic secrecy rate of $U_{m}$. At high SNR regimes increasing $P_{BS}$ has no effect on the ergodic secrecy rate of $U_{m}$, because at high SNR regimes the ergodic secrecy rate of $U_{m}$ only depends on $|\tilde{g}_{m}|^{2}$, $|\tilde{g}_{n}|^{2}$ and $|\tilde{g}_{e}|^{2}$. Moreover, it is shown  that existence of the external eavesdropper degrades the ergodic secrecy rate of $U_{m}$.

Fig. 6 illustrates the ergodic rate and ergodic secrecy rate of $U_{n}$ and $U_{m}$  versus $a_{m}^{2}$ with and without $U_{e}$. It is shown that increasing $a_{m}^{2}$ decreases the ergodic secrecy rate of $U_{m}$ and decreases the ergodic rate/ ergodic secrecy rate of $U_{n}$ due to increment of  the allocated power of  $U_{m}$ and decrement of the  allocated power of $U_{n}$.

\section{Conclusion}
In this paper, we studied the secrecy performance of a NOMA system, while the far user acts as an internal  eavesdropper with and without the external eavesdropper. We presented the closed-forms for the secrecy outage probability and ergodic secrecy  rate of the users. Investigating the optimized power allocation coefficients in order to minimize the secrecy outage probability of the system and also the ergodic secrecy rates of the users is a subject of our future works.
\vspace*{-.5em}

\end{document}